\documentclass[aps,prl,twocolumn,showpacs,superscriptaddress]{revtex4-1}
\usepackage{graphicx}
\usepackage{amsmath}
\usepackage{textcomp}
\usepackage{tikz}
\newcommand\MyTextBox[2]{%
\begin{tikzpicture}[remember picture,overlay]
\node[anchor=north,draw,yshift=-#1,text width=18cm] at (current page.north)
{\parbox[t][1.5cm][c]{\linewidth}{#2}};
\end{tikzpicture}%
}

\begin{document}

\title{Unveiling the optical properties of a metamaterial synthesized by electron-beam-induced deposition}
\author{P. Wo{\'z}niak}
\email{pawel.wozniak@mpl.mpg.de}
\thanks{P. Wo{\'z}niak and K. H{\"o}flich contributed equally to this work.}
\affiliation{Max Planck Institute for the Science of Light, G\"unther-Scharowsky-Str.1, D-91058 Erlangen, Germany}
\affiliation{Institute of Optics, Information and Photonics, Friedrich-Alexander-University Erlangen-Nuremberg, Staudtstr. 7/B2, D-91058 Erlangen, Germany}
\author{K. H{\"o}flich} 
\thanks{P. Wo{\'z}niak and K. H{\"o}flich contributed equally to this work.}
\affiliation{Institute of Nanoarchitectures for Energy Conversion, Helmholtz Centre Berlin for Materials and Energy, Hahn-Meitner-Platz 1, D-14109 Berlin}
\affiliation{Max Planck Institute for the Science of Light, G\"unther-Scharowsky-Str.1, D-91058 Erlangen, Germany}
\author{G. Br{\"o}nstrup} 
\affiliation{Max Planck Institute for the Science of Light, G\"unther-Scharowsky-Str.1, D-91058 Erlangen, Germany}
\affiliation{Institute of Nanoarchitectures for Energy Conversion, Helmholtz Centre Berlin for Materials and Energy, Hahn-Meitner-Platz 1, D-14109 Berlin}
\author{P. Banzer}
\affiliation{Max Planck Institute for the Science of Light, G\"unther-Scharowsky-Str.1, D-91058 Erlangen, Germany}
\affiliation{Institute of Optics, Information and Photonics, Friedrich-Alexander-University Erlangen-Nuremberg, Staudtstr. 7/B2, D-91058 Erlangen, Germany}
\author{S. Christiansen}
\affiliation{Institute of Nanoarchitectures for Energy Conversion, Helmholtz Centre Berlin for Materials and Energy, Hahn-Meitner-Platz 1, D-14109 Berlin}
\affiliation{Max Planck Institute for the Science of Light, G\"unther-Scharowsky-Str.1, D-91058 Erlangen, Germany}
\author{G. Leuchs}
\affiliation{Max Planck Institute for the Science of Light, G\"unther-Scharowsky-Str.1, D-91058 Erlangen, Germany}
\affiliation{Institute of Optics, Information and Photonics, Friedrich-Alexander-University Erlangen-Nuremberg, Staudtstr. 7/B2, D-91058 Erlangen, Germany}
\date{\today}

\MyTextBox{+0.5cm}{This is an author-created, un-copyedited version of an article accepted for publication/published in Nanotechnology. IOP Publishing Ltd is not responsible for any errors or omissions in this version of the manuscript or any version derived from it. The Version of Record is available online at DOI: 10.1088/0957-4484/27/2/025705.}

\begin{abstract}
The direct writing using a focused electron beam allows for fabricating truly three-dimensional structures of sub-wavelength dimensions in the visible spectral regime. The resulting sophisticated geometries are perfectly suited for studying light-matter interaction at the nanoscale. Their overall optical response will strongly depend not only on geometry but also on the optical properties of the deposited material. In case of the typically used metal-organic precursors, the deposits show a substructure of metallic nanocrystals embedded in a carbonaceous matrix. Since gold-containing precursor media are especially interesting for optical applications, we experimentally determine the effective permittivity of such an effective material. Our experiment is based on spectroscopic measurements of planar deposits. The retrieved permittivity shows a systematic dependence on the gold particle density and cannot be sufficiently described using the common Maxwell-Garnett approach for effective medium.
 
\end{abstract}
\maketitle
\section{Introduction}
Nowadays, the fast development of nanofabrication methods provides access to functional structures with geometries of sub-wavelength dimensions even in the visible regime. Thereby, fundamental questions of light-matter interaction can be addressed as done within the intensely studied field of nano-optics. Especially, nanostructures made of materials having a free electron gas \cite{Boltasseva2011}, excitable to collective oscillations by light (plasmon-polaritons), provide the possibility of tailored light manipulation and concentration \cite{Banzer2010,Schuller2010,Baffou2014}. Such sub-wavelength structures are key for the design of metamaterials \cite{Cai2010}. However, the fabrication of truly three-dimensional nanostructures still represents a significant challenge. While, for example, direct laser writing provides three-dimensional and purely metallic \cite{Gansel2009} as well as dielectric \cite{Thiel2010} structures, it is limited primarily not in terms of optical resolution, but most possibly due to diffusion constraints during the electroless plating. In contrast, direct writing using a focused electron beam is capable of delivering three-dimensional features, smaller than tens of nanometer in a highly flexible and precise single-step process \cite{Utke2012, Huth2012,Hoeflich2011}. The electron beam induced deposition (EBID) process is based on local decomposition of inserted precursor-gas molecules by an electron beam within a vacuum chamber of a scanning electron microscope. Since typical precursors are of metal-organic type, not only their metallic constituents remain after deposition, although this would be the ideal case. Instead, a substantial amount of ligand components as well as of elements of the residual gas in the vacuum chamber (carbon, oxygen and hydrogen) are incorporated \cite{Utke2012,Botman2009}. Thus, the resulting composite can be viewed as a metamaterial or an effective material by itself, consisting of single-crystalline nanoparticles (e.g. gold) embedded into a carbonaceous matrix \cite{Hoeflich2011}. While the electric properties of granular materials such as the EBID material are well understood \cite{Beloborodov2007} and proven to be promising for sensing applications \cite{Huth2014,Kolb2013}, a complete optical description of EBID metamaterials based on a combined experimental and theoretical study still has not been reported yet. This is caused by the relatively small areas which can be conventionally fabricated with the EBID process, not allowing for optical characterization with standard methods such as ellipsometry. First attempts using micro-ellipsometry showed a behaviour close to a Maxwell-Garnett effective medium for a gold containing EBID material \cite{Utke2011}. Recently, the Maxwell-Garnett theory was also used to describe material properties of helices made of platinum-EBID composite \cite{Esposito2014} to numerically study their circular dichroism. While the overall agreement between experiment and numerical simulations was reasonable, distinct deviations became obvious for both the long wavelength range and the actual strength of the observed optical activity \cite{Esposito2014}. This trend was also emphasized in a recent numerical study of similar composite materials \cite{Etruch2014}.\\
Here, we present the results of spectrometric measurements in combination with a numerical algorithm based on the transfer-matrix approach for retrieval of the permittivity of gold-EBID 10x10 \textmu m$^2$ pad-like deposits (see Fig. \ref{fig_setup}c). The investigation concerns deposits of different thicknesses to systematically study the dependence of the permittivity on the density of the nanoinclusions in the carbonaceous matrix. In addition, the comparison of analytical results based on Mie theory with an effective medium model such as Maxwell-
Garnett proves that although the dipole resonance of the gold particles dominantly contributes to the optical response of the EBID composite, the conventional Maxwell-Garnett approach cannot be employed to describe the effective optical properties of the composite. The presented experimentally retrieved permittivity is expected to give better understanding of optical performance of nanostructures fabricated via EBID, paving the way for sophisticated nano-optic applications. 
\section{Sample Preparation}
\begin{figure}
\includegraphics[scale=1.3]{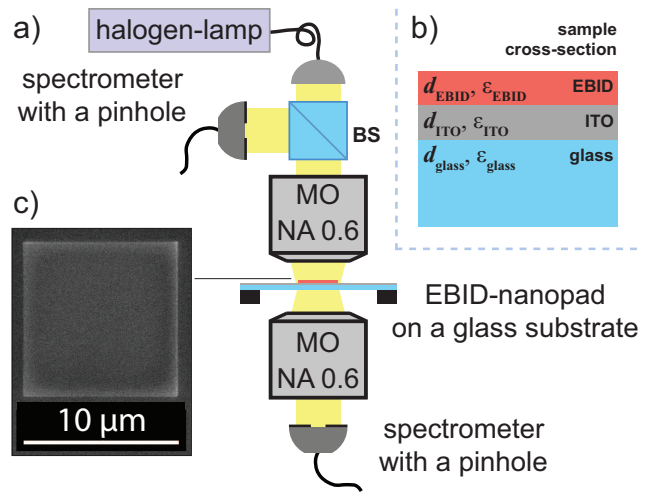}
\caption{\footnotesize a) Microscopy setup for spectral measurements of EBID nanopads. The sample is illuminated with white light emitted by a halogen-lamp. A beam splitter (BS) between the lamp and the first microscope objective (MO) guides part of the reflected light onto a detector. The  central part of the nanopad is imaged onto a pin hole with diameter of 600 \textmu m placed in front of the detectors. b) Cross-section through the nanopad and the substrate (layer thickness are not drawn to scale). c) Scanning electron micrograph of an investigated EBID pad.}
\label{fig_setup}
\end{figure}

The experimental investigation started with the fabrication of a substrate which consists of a 170 \textmu m thick glass (BK7) plate with a conductive thin layer of indium tin oxide (ITO) on top \footnote{Commercial ITO-on-glass substrate was avoided due to undefined thickness of the ITO coating and its pronounced crystallinity causing strong scattering effects.}. Both materials were characterized using ellipsometry and show typical properties \cite{IS}. The same ellipsometry measurement was used to determine most accurately the thickness of the ITO coating (36 nm). \\
The EBID process itself was carried out in a vacuum chamber of a dual-beam instrument (FEI Strata DB 235) equipped with a gas-injection system for dimethyl-gold(III)-acetyl-acetonate (Me$_{2}$Au(acac)) as the precursor gas \cite{Wnuk2010}. The injected precursor molecules adsorb, diffuse and desorb onto the sample surface where they are locally cracked by the focused electron beam \cite{Utke2010}. Thereby, the EBID technique does not only allow for the realization of complex three-dimensional geometries \cite{Hoeflich2012, Esposito2014} but also for depositing them on even non-planar substrates, if these are at least weakly conductive \cite{Hoeflich2011}. During the molecule dissociation, their non-volatile parts form the deposit while the volatile constituents are pumped out \cite{Wnuk2010}. \\
The deposition process depends on the energy and current of the electron-beam, on the dwell time (the time the beam resides at each position) as well as on the scanning raster. All these factors were previously optimized for fabrication of high-resolution three-dimensional nanostructures \cite{Hoeflich2011}. The aim in the current work was to keep the deposition parameters of the pad deposits as close as possible near the range used for nanostructures to unveil their material properties. Correspondingly, the retrieved permittivity reliably describes the material of nanostructures written by EBID. Since the optimized parameters would lead to unrealistically long deposition time of such deposits, the dwell time was drastically reduced down to 1 \textmu s and the lateral sizes were restricted to 10x10 \textmu m$^2$ while maintaining the other parameters (see scanning electron micrograph of a typical EBID pad in Fig.~\ref{fig_setup}c). For the systematic investigation, EBID deposits with thicknesses ranging from 15 nm up to 50 nm (far below below Fabry-P{\'e}rot cavity condition) were fabricated and topographically characterized by atomic force microscopy (AFM). The AFM scans confirmed that at the edges, all pads have increased thickness with respect to the perfectly smooth and flat interior. This is caused by low pressure of the precursor gas \cite{Wnuk2010}, resulting in a mass-transport-limited (also called molecule-limited) deposition regime \cite{Szkudlarek2014, Winkler2014,Plank2013}. In this regime all available precursor molecules within the square deposition region are consumed and dissociated by the primary and the secondary electrons. This leads to a decreased deposition rate in the central area as well as to stronger co-deposition of carbon from the residual chamber gases \cite{Utke2010,Bernau2010}. The diffusion of the intact precursor molecules outside the irradiated region causes an increased precursor supply at the edges leading to enhanced deposition there \cite{Winkler2014}. Hence, each fabricated pad exhibits a 9x9 \textmu m$^2$ large surface area of high quality and constant thickness which were utilized for spectroscopic measurements in an optical microscopic setup.\\
\section{Methods}
To measure optical spectra, a home-build microscopic setup was used (see Fig.~\ref{fig_setup}a). A halogen lamp emitted unpolarized and noncoherent light within the spectral range of interest between 480 and 900 nm. The sample was illuminated using a 60x microscope objective which also collected the reflected light. For separation of the incident and the reflected light, a beam splitter was used between the source and the objective. The position of the microscope objective relative to the sample surface was carefully adjusted to obtain plane-wave illumination at normal incidence and to image the sample surface onto the detector plane. The area of interest could be selected using a pinhole in the image plane. The size of this additional aperture was chosen to be 600 \textmu m in diameter guaranteeing the measurement of the signal from the flat inner part of the deposit solely, while maintaining a sufficient signal-to-noise ratio. Each EBID pad was measured separately by placing it on the optical axis such that the propagation of the incoming plane-wave was normal with regard to the sample surface. In the same way, the light transmitted through the sample was measured. \\
The experimental scheme described above permits interpreting the measurement data as spectra of reflected ($I_{\text{R}}$) and transmitted ($I_{\text{T}}$) intensities of an incoming plane-wave ($I_{\text{0}}$) by a multilayer system (see Fig.~\ref{fig_setup}b). Such a structure can be analytically described using a transfer matrix \cite{IS}: 
\begin{equation}
\begin{bmatrix}
I_{\text{R}} \\
I_{\text{T}} 
\end{bmatrix}
=
\begin{bmatrix}
R_{\text{EBID}}R_{\text{ITO}}R_{\text{glass}}\\
T_{\text{EBID}}T_{\text{ITO}}T_{\text{glass}}
\end{bmatrix} I_{\text{0}}\text{,}
\label{eq_t_matrix}
\end{equation}
\begin{figure*}
\includegraphics[scale=0.40]{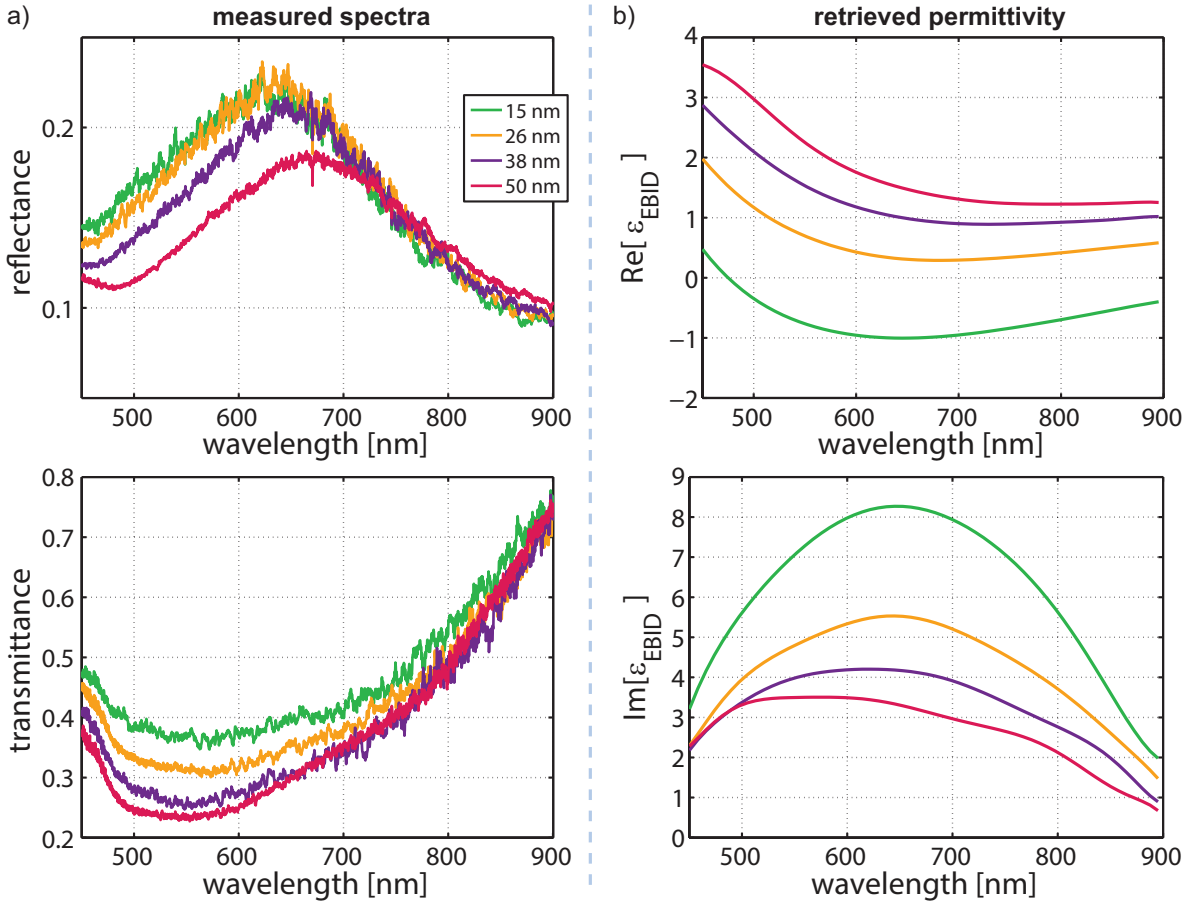}
\caption{\footnotesize a) Experimental reflectance and transmittance spectra of EBID nanopads of thicknesses between 15 and 50 nm. b) Retrieved permittivity as a function of the pad thickness; for the sake of clarity, the measurement noise is removed by fitting the retrieved real and imaginary parts of $\epsilon_{\text{EBID}}$($\lambda$) with a polynomial functions. The green curve of the thinnest deposit is expected to approximate the best  material properties of nanostructures, fabricated with EBID.}
\label{fig_spectra}
\end{figure*}
where $R$ and $T$ are reflection and transmission matrices describing propagation and multiple reflection of light within the EBID layer, the ITO layer and the glass substrate, respectively. Each matrix contains information about the geometrical (thickness $d$) and optical (permittivity $\epsilon$) properties of the corresponding layer. The only unknowns in the equation above are the real and imaginary part of $\epsilon_{\text{EBID}}$. All further parameters describing the ITO-on-glass substrate as well as the EBID layer thicknesses are determined experimentally. Consequently, the optical properties of the EBID material can be retrieved from measurements of reflection and transmission. Due to the sub-wavelength thickness, the light within the EBID and the ITO layers ($R_{\text{EBID}}\text{, }R_{\text{ITO}}\text{, }T_{\text{EBID}}\text{ and }T_{\text{ITO}}$) must be described as coherent \cite{Moharam1995} in the investigated spectral range. In contrast, the glass substrate is two orders of magnitude thicker than $\lambda$. Therefore and due to the used light source, light propagation across the entire sample requires a semi-coherent description \cite{Harbecke1986}. Furthermore, according to Eq.~\ref{eq_t_matrix} the measured values $I_{\text{R}}$ and $I_{\text{T}}$ have to be normalized with respect to the incident intensity $I_{\text{I}}$. To this end, the reflection from and the transmission through the ITO-on-glass substrate was measured and the theoretical reflectance and transmittance coefficients were calculated \cite{IS}.\\
Though the calculation of the sample's reflectance and transmittance is straight forward if the complex permittivity of the EBID layer is known, the inverse problem is not. In addition, there are several distinct variation of $\epsilon_{\text{EBID}}$ which result in the same reflectance and transmittance of the sample. Therefore, this inverse problem was tackled using a brute force search, i.e., calculating the reflectance and transmittance of the sample for several thousand possible complex values of $\epsilon_{\text{EBID}}$ for each wavelength and comparing them the experimental results. Thereby, all local minima in the two-dimensional permittivity-space corresponding the respective smallest deviation from the measured spectra were identified. By subsequent refining of the region around the minima several steady solutions of $\epsilon_{\text{EBID}}$($\lambda$) were found. However, there is only one physical solution which has to be determined. Accordingly, all dispersions which exhibits a constant increase of the real part of the refractive index ($d\text{Re}[n]/d\lambda=const$ where $n=\sqrt{\epsilon}$) in the whole investigated spectral range are excluded. Such solutions have unrealistically high values and appear repeatedly while extending the plane of possible solutions \cite{IS}. In contrast, the only dispersion which does not follow this tendency, has values in a reasonable range and meaningfully reflects the inner structure of the EBID composite.
\section{Results and Discussion}
\begin{figure}
\includegraphics[scale=0.40]{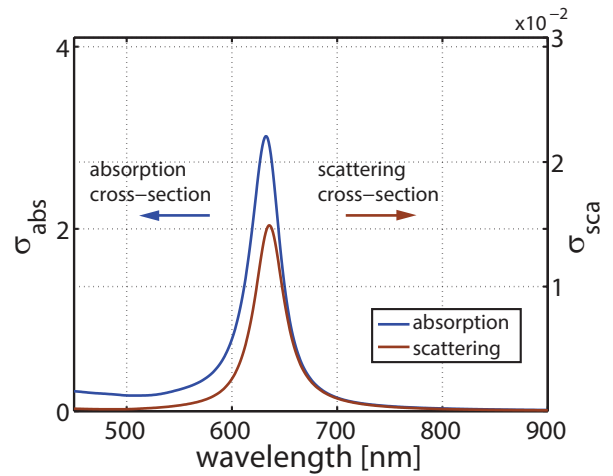}
\caption{\footnotesize Mie scattering and absorption cross-sections of isolated gold particles embedded in diamond environment under plane-wave excitation. The particles have particles the mean diameter  of 4 nm with a lognormal size distribution of 50$\%$ standard deviation.}
\label{fig_mie}
\end{figure}

\begin{figure}
\includegraphics[scale=0.40]{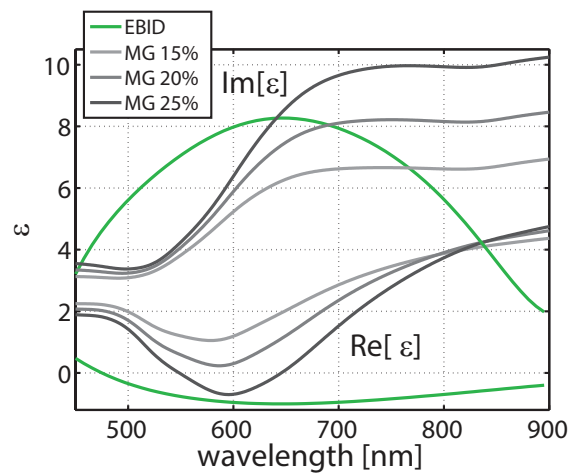}
\caption{\footnotesize Comparison of the retrieved permittivity of the EBID material (green) with the Maxwell-Garnett effective medium theory for three different volume filling fractions of 15$\%$, 20$\%$ and 25$\%$.}
\label{fig_mg}
\end{figure}

Fig.~\ref{fig_spectra}a shows the experimental reflectance and transmittance data of the EBID-ITO-glass system for different deposit thicknesses. All reflectance spectra exhibit distributions with a maximum value observed at around 620 nm, which slightly red-shifts for increasing layer thickness. Likewise, the transmittance spectra feature minima at around 550 nm and also show the expected decrease in transmittance for thicker pads. The corresponding real and imaginary parts of the retrieved permittivity are presented in Fig.~\ref{fig_spectra}b (for the sake of clarity, the measurement noise is removed by fitting the retrieved real and imaginary parts of $\epsilon_{\text{EBID}}$($\lambda$) with polynomial functions) and show a clear systematic dependence on the thickness of the EBID pads. The real part of $\epsilon_{\text{EBID}}$ decreases when decreasing deposit thickness, approaching even metallic behaviour for the thinnest pad. In contrast, the imaginary part increases with decreasing thickness and moves to form a broad peak centered at around 650 nm. This tendency implies that the EBID composite behaves more and more dielectric for thicker deposits. \\
This observation is consistent with a chemisorption of the gold-containing precursor molecules onto the sample surface together with possibly decreasing vapor pressure over the long deposition times \cite{Utke2010, Bernau2010}. As mentioned, the concave shape of the pads implies a mass-transport-limited deposition regime \cite{Winkler2014} in which excess electrons start to dissociate residual gases mainly hydrocarbons present in the vacuum chamber. For typical vacuum pressures around $10^{-6}$ hPa, still a large number of molecules sufficient to form one monolayer per second impinges onto the substrate only from the residual gases \cite{Utke2010} and and possibly due to a longer resident time of some precursor-gas ligands at the irradiated spot \cite{Szkudlarek2014}. Thereby, after the depletion of the chemisorbed initial layer the codeposition of carbon, oxygen and hydrogen becomes dominant and increases the relative carbon content in the thicker pads. For long deposition time, this effect becomes particularly important if the precursor vapor pressure decreases with time. Thus, the carbon content of the deposits rises along with the deposition time and with the deposit thickness \cite{Bernau2010}. As a consequence, properties of the EBID composite can be tuned by the density of the gold inclusions, from strongly dielectric to slightly metallic optical behavior. \\
In that respect, the permittivity of the thinnest pad is expected to correspond to the material properties of nanostructures which can be fabricated using EBID. Fine structures (e.g. pillars or needles) are fabricated optimally with a dwell time of 720 \textmu s, initiating local heating, while staying in the mass-transport-limited deposition regime. Furthermore, the slender geometry, together with the overall short deposition time, supports the heating effect, which ensures a relatively high content of gold around 28 at$\%$ \cite{Hoeflich2011}. \\
Concerning the spectral characteristic of the retrieved permittivity, the prominent resonance in $\text{Im}[\epsilon_{\text{EBID}}]$ at around 650 nm can be attributed to the dipole resonance of the gold nanoparticles within the carbonaceous matrix. The mean particle diameter of the singly crystalline particles varies around 4 nm ranging approximately from 2 nm to 7 - 8 nm with decreasing number of larger particles \cite{Hoeflich2011, Riazanova2012}.  Correspondingly, Fig.~\ref{fig_mie} shows Mie cross-sections of absorption (blue line) and of scattering (brown line) for isolated spherical gold particles of a mean diameter of 4 nm with a lognormal size distribution of 50$\%$ standard deviation under plane-wave illumination \cite{IS}. The EBID matrix is approximated by non-absorbing diamond \cite{Peter1923} (solid line) while gold properties are taken from \cite{Johnson1972}. As expected, the Mie cross-sections are dominated by an absorptive dipole resonance around 630 nm which corresponds to the absorption properties retrieved for the EBID matrix. In this size regime the shift of the resonance position due to the nanoparticle size distribution is negligible. Thus, the Mie  resonance is comparatively narrow indicating that the broadening of the metamaterial resonance is due to further loss mechanisms besides the effect of the size distribution and the intrinsic absorption of the gold inclusions. In that respect, absorption of the carbonaceous matrix as well as interaction between the particles, which are not included in the classical Mie approach, play a significant role. Careful analytical considerations \cite{Quinten1996} show that, tiny losses of the matrix will not only slightly red-shift the peak position, but also cause strong broadening of the resonance. Therefore, losses of the carbonaceous matrix, especially for the thicker deposits, are responsible for the very broad resonance of in $\text{Im}[\epsilon_{\text{EBID}}]$. \\  
Based on these results, description of $\epsilon_{\text{EBID}}$($\lambda$) using an effective medium approach based on dipoles in a dielectric environment seems to be promising. Considering the inner structure of the EBID material (randomly distributed metallic nanocrystals of deep sub-wavelength dimensions within a dielectric matrix), the Maxwell-Garnett (MG) theory \cite{Quinten2011} constitutes the obvious choice. For that purpose the matrix material is described as amorphous carbon \cite{Schnaiter1998} to account for the matrix loss and gold properties are taken again from \cite{Johnson1972}. Since the actual value of the volume filling factor of gold in the composite is not known, it can therefore act as tuning factor. Fig.~\ref{fig_mg} shows comparison between the retrieved permittivity of the thinnest pad and MG effective media for three different values of filling factor: 15$\%$, 20$\%$, 25$\%$. None of the possible combinations can reproduce the retrieved $\epsilon_{\text{EBID}}$($\lambda$). Neither tuning of the filling factor nor substitution of the material properties of gold and the carbonaceous matrix (including absorption) from other material databases significantly changes the output of the MG calculations. This indicates that the EBID composite does not meet MG conditions \cite{Quinten2011,Etruch2014}. Indeed, the particles can be safely described within the quasistatic approximation, the absorptive dipole-type resonance fits well to the Maxwell Garnett approach considering pure dipole-dipole interaction under the assumption of negligible scattering. However, well separation of the particles, required by MG cannot be taken for granted. The close spacing between the gold inclusions will presumably cause both not only near-field interactions \cite{Etruch2014} but also local electron tunnelling through the carbonaceous matrix. As proven for the electrical applications of EBID materials, the gold nanocrystals are not electrically isolated within the carbonaceous matrix, but instead, electron transport, due to thermally assisted tunneling of electrons, takes place \cite{Huth2014}. Furthermore, the mass-limited deposition regime and hence the varying amount of co-deposited carbon influences the local particle density. An appropriate choice for the  extension of Maxwell-Garnett seems to be the introduction of self-consistency, and thereby the iterative correction of the properties of the carbonaceous matrix by the effective permittivity itself \cite{Polder1946}. In addition, the formation effect of differently shaped particle aggregates and their influence of higher order multipolar contributions need to be accounted as well \cite{Granqvist1977,Etruch2014}. However, considering all these effects together and due to the lack of knowledge concerning the optical properties of the carbonaceous matrix, a consistent effective material approach relying on only few and physically meaningful parameters still represents a significant challenge.
\section{Conclusions} 
In summary, the dielectric function of an EBID material consisting of single-crystalline gold particles dispersed in a carbonaceous matrix is studied experimentally. The retrieved effective permittivities show a systematic dependence on the layer thickness and, thus, the particle density within the deposit. The effective permittivity for the highest metal content represents
the best approximation of the material properties of EBID nanostructures. While the prominent absorption feature present in the retrieved imaginary parts of the EBID permittivities could be attributed to the dipole resonances of the embedded gold nanoparticles via Mie calculations, the optical properties of the material cannot be described by using a standard Maxwell-Garnett approach. Remarkably, this material also shows a very interesting optical behaviour at wavelengths around 450 nm where the real part of its permittivity exhibits a zero-crossing, which might require more attention. While the presented study is based on Me$_{2}$Au(acac) as an exemplary gas precursor, the discussed method can be applied to any type of EBID-based metamaterial fabricated using metal-organic precursors. Therewith, the numerical and experimental study of complex nanostructures made of EBID metamaterials can be envisaged.\\
\subsection{Acknowledgements}
The research leading to these results received funding from the Helmholtz Association within the Helmholtz Postdoc Programme as well as from the European Union Seventh Framework Programme under Grant Agreement No. 280566 (\textit{www.univsem.eu}) and under Grant Agreement No. 258868 (\textit{www.lcaos.eu}). \\

\bibliographystyle{unsrt}
\bibliography{biblioteca}
\end{document}


\title{Unveiling the optical properties of a metamaterial synthesized by electron-beam-induced deposition \\
SUPPLEMENTARY INFORMATION}
\author{P. Wo{\'z}niak}
\email{pawel.wozniak@mpl.mpg.de}
\thanks{P. Wo{\'z}niak and K. H{\"o}flich contributed equally to this work.}
\affiliation{Max Planck Institute for the Science of Light, G\"unther-Scharowsky-Str.1, D-91058 Erlangen, Germany}
\affiliation{Institute of Optics, Information and Photonics, Friedrich-Alexander-University Erlangen-Nuremberg, Staudtstr. 7/B2, D-91058 Erlangen, Germany}
\author{K. H{\"o}flich}
\thanks{P. Wo{\'z}niak and K. H{\"o}flich contributed equally to this work.}
\affiliation{Helmholtz Centre Berlin for Materials and Energy, Institute of Nanoarchitectures for Energy Conversion, Hahn-Meitner-Platz 1, D--14109 Berlin }
\affiliation{Max Planck Institute for the Science of Light, G\"unther-Scharowsky-Str.1, D-91058 Erlangen, Germany}
\author{G. Br{\"o}nstrup} 
\affiliation{Max Planck Institute for the Science of Light, G\"unther-Scharowsky-Str.1, D-91058 Erlangen, Germany}
\affiliation{Helmholtz Centre Berlin for Materials and Energy, Institute of Nanoarchitectures for Energy Conversion, Hahn-Meitner-Platz 1, D--14109 Berlin }
\author{P. Banzer}
\affiliation{Max Planck Institute for the Science of Light, G\"unther-Scharowsky-Str.1, D-91058 Erlangen, Germany}
\affiliation{Institute of Optics, Information and Photonics, Friedrich-Alexander-University Erlangen-Nuremberg, Staudtstr. 7/B2, D-91058 Erlangen, Germany}
\author{S. Christiansen}
\affiliation{Helmholtz Centre Berlin for Materials and Energy, Institute of Nanoarchitectures for Energy Conversion, Hahn-Meitner-Platz 1, D--14109 Berlin }
\affiliation{Max Planck Institute for the Science of Light, G\"unther-Scharowsky-Str.1, D-91058 Erlangen, Germany}
\author{G. Leuchs}
\affiliation{Max Planck Institute for the Science of Light, G\"unther-Scharowsky-Str.1, D-91058 Erlangen, Germany}
\affiliation{Institute of Optics, Information and Photonics, Friedrich-Alexander-University Erlangen-Nuremberg, Staudtstr. 7/B2, D-91058 Erlangen, Germany}
\date{\today}

\MyTextBox{+0.5cm}{This is an author-created, un-copyedited version of an article accepted for publication/published in Nanotechnology. IOP Publishing Ltd is not responsible for any errors or omissions in this version of the manuscript or any version derived from it. The Version of Record is available online at DOI: 10.1088/0957-4484/27/2/025705.}
 
\maketitle

\subsection{Stable Implementation of Layer System and Analysis of Multiple Solutions}
\begin{figure}
\includegraphics[scale=0.50]{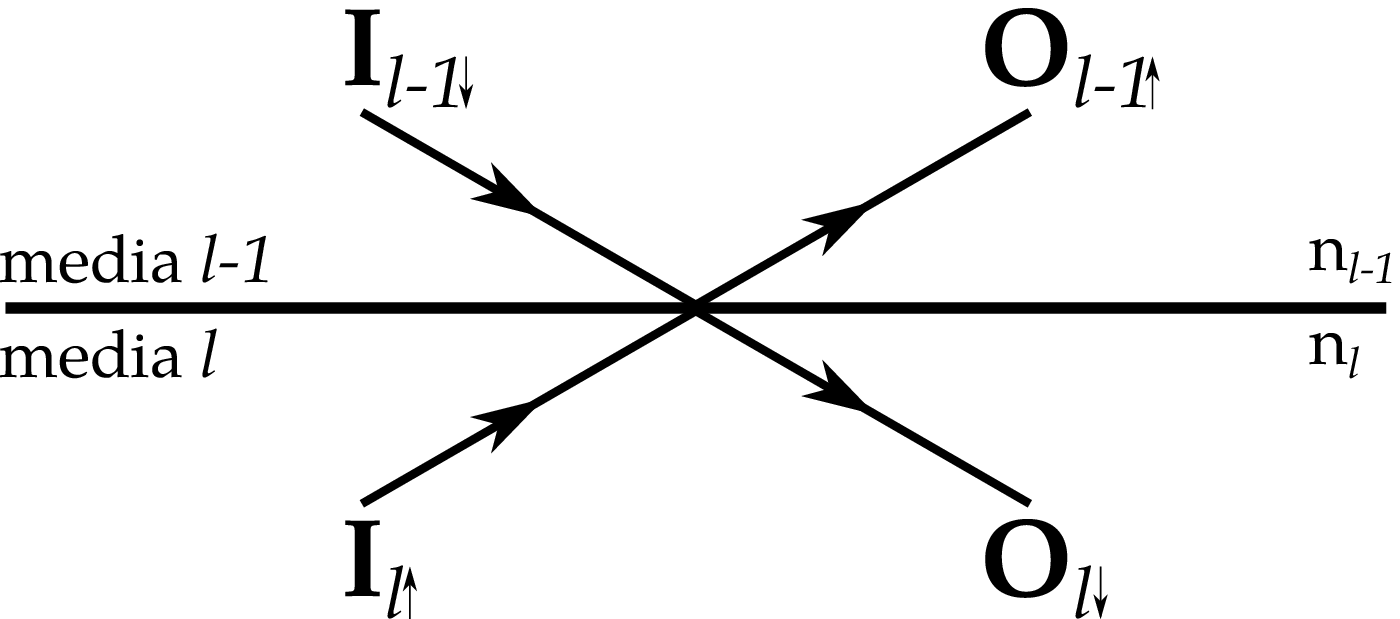}
\caption{\footnotesize Incoming (\textbf{I}) and outgoing (\textbf{O}) waves at an interface between the media $l-1$ and $l$ due to multiple reflection. }
\label{fig_interface}
\end{figure}

\begin{figure}
\includegraphics[scale=0.50]{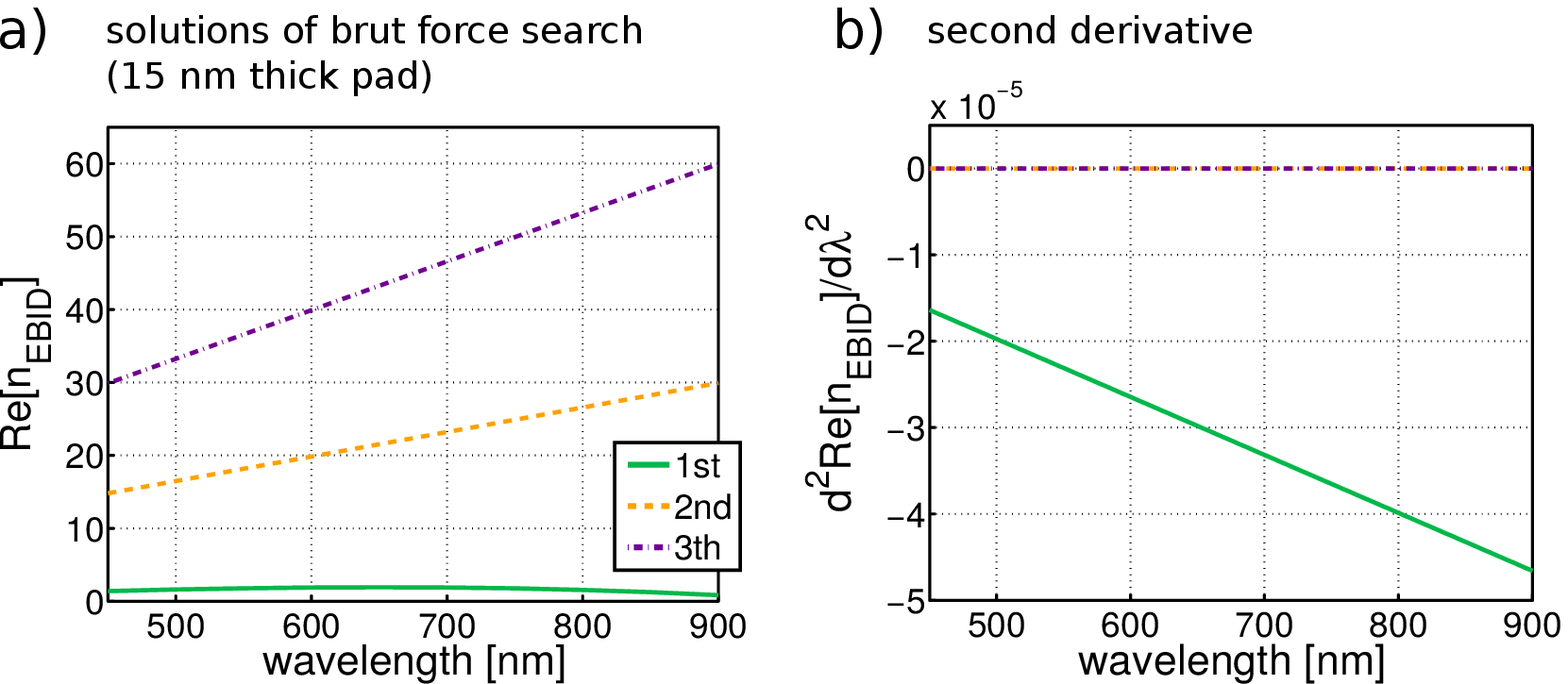}
\caption{\footnotesize Multiple solutions of the brute force search algorithm for the EBID pad of 15 nm: real part of the refractive indexes (a) and corresponding second derivatives (b).}
\label{fig_multiple_solutions}
\end{figure}

As mentioned in the manuscript, the sample can be considered as a layer system on a thick glass substrate (see Fig. 1b). Due to the presence of many interfaces, the light propagating through the system will undergo multiple reflections. Moreover, because of the sub-wavelength thickness of the EBID and the ITO layers, light within these layers must be considered as coherent even for the used light source. Such a situation can be described with a transfer matrix based on field continuity at each interface. Supplementary Fig.~\ref{fig_interface} shows an interface between media $l-1$ and $l$, where $\textbf{I}_{l-1\downarrow}$ and $\textbf{I}_{l\uparrow}$ represent the electric field of the plane-wave incoming from above and below respectively, while $\textbf{O}_{l-1\uparrow}$ and $\textbf{O}_{l\downarrow}$ represent the corresponding outgoing fields. Consequently, propagating $\textbf{O}_{l\downarrow}$ becomes $\textbf{I}_{l\downarrow}$ at the next interface and analogically $\textbf{I}_{l\uparrow}$ can be interpreted as $\textbf{O}_{l\uparrow}$, which propagated from the interface $l/l+1$ through the media $l$. The relation between \textbf{I}  and \textbf{O} is following:
\begin{align}
\textbf{O}_{l\downarrow}&=\textbf{I}_{l\downarrow}e^{-i\delta_{l}} \nonumber \\
\textbf{I}_{l\uparrow}&=\textbf{O}_{l\uparrow}e^{+i\delta_{l}},
\end{align}
where
\begin{align}
\delta_{l}&=\text{k}_{0}\text{n}_{l}\gamma_{l}\text{d}_{l} \nonumber
\end{align}
is the phase that the field gains while propagating from one interface to another. In the equation above, $\text{k}_{0}$ is the wave-vector, $\text{n}_{l}$ the refractive index (complex for lossy media) of the $l$-th media and $\text{d}_{l}$ is the thickness of the $l$-th layer. The factor $\gamma_{l}$ is the directional cosine in the $l$-th media and it is equal to 
$[\text{n}_{l}^{2}-(\text{n}_{l-1}sin\theta_{l-1})^{2}]^{1/2}$ . By Snell's law, it can be rewritten as $\gamma_{l}=[\text{n}_{l}^{2}-(\text{n}_{\text{in}}sin\theta_{\text{in}})^{2}]^{1/2}$, where $\text{n}_{\text{in}}$ is the refractive index of the input media and $\theta_{\text{in}}$ the incidence angle of the incoming plane-wave. \\
If the product $\gamma_{l}\text{d}_{l}$ is large and positive (lossy media or evanescent fields), it will cause numerical instabilities while inverting the Eq. 1 to retrieve the dielectric function of an EBID-pad. Therefore, a stabilization technique was implemented as proposed by Moharam et al. \cite{Moharam1995}. \\
The light propagating in the 170 \textmu m thick glass substrate also undergoes multiple reflections. However, the thickness of the substrate is two orders of magnitude larger than $\lambda$ (between 480 and 900 nm) and allows for infinite summation over intensities instead of fields. Thus, light propagation across the entire sample can be described using a semi-coherent formalism \cite{Harbecke1986}. Accordingly, the multiple reflected waves in the $\text{R}_{\text{EBID}}$, $\text{R}_{\text{ITO}}$, $\text{T}_{\text{EBID}}$ and $\text{T}_{\text{ITO}}$ matrices need to meet the boundary conditions (continuity of the tangential electromagnetic field), required by Maxwell's equations at an interface, while in the $\text{R}_{\text{glass}}$ and $\text{T}_{\text{glass}}$ the waves are summed up with the Fresnel reflectance and transmittance coefficients.
Inversion of the problem, concisely written with Eq. 1, with respect to the dielectric function of the EBID composite cannot be solved analytically. Therefore, a brute force search for local minima in the plane of the real and the imaginary parts of $\epsilon_{\text{EBID}}$($\lambda$) was applied. Due to non-linear equations related to the light propagation across the sample, multiple solutions are expected to be found. From a mathematical point of view all roots are correct. However, there is only one solution having physical meaning. \\
As an example, supplementary Fig.~\ref{fig_multiple_solutions}a shows the real part of the found refractive indices ($\text{Re}[n_{\text{EBID}}]=\text{Re}[\epsilon_{\text{EBID}}^{1/2}]$) for the pad of thickness of 15 nm. Apart from the first solution (green solid curve), two others were found in the presented range of $\text{Re}[n_{\text{EBID}}]$ (yellow and violet dash curves). These solutions have unrealistically high values (given the constituent elements of the composite - gold and carbon with faint amount of oxygen and hydrogen) and they exhibit a strictly linear dispersion with respect to $\lambda$. Supplementary Fig 2b shows the corresponding derivatives (due to different slope of the second and the third curves, the second derivatives are plotted for convenience). Further extending the plane of possible solutions revealed that solutions of $dRe[n]/d\lambda=const$ appear repeatedly, increasing the value of $\text{Re}[n_{\text{EBID}}]$ towards infinity. The only solution not following this trend is the first one (green curve, corresponding to the the green curves in Fig. 2), which can be meaningfully explained by the structure of the composite as presented in the manuscript. 
Analogous solutions were found for the other investigated EBID pads. 
\subsection{Mie calculations}
The scattering problem of planar waves incident onto a spherical obstacle can be solved analytically based on the solution found by Gustav Mie \cite{Mie1908}.
For the Mie calculations presented in the manuscript a free software package by Philip Laven (\textit{http://www.philiplaven.com/mieplot.htm}) was used. It allows for the calculation of Mie scattering efficiencies versus wavelength and for the implementation of wavelength-dependent material properties for the scattering sphere as well as for the embedding medium. As mentioned in the manuscript, the Mie calculations solve the problem for a single sphere only and therefore include no interaction between the particles. Although this assumption is not valid in the case of the EBID material in which the scatterers are closely spaced, the mainly absorptive Mie resonance reproduces the resonant behavior of the imaginary part of the effective permittivity very well. In that respect one has to notice that the imaginary part of the permittivity reflects the relative phase delay of oscillations of charges within the material caused by the electric field of the incident wave and thereby includes all loss mechanisms, while the real part describes the amplitude of the oscillation. 
\begin{figure}
\includegraphics[width = \linewidth]{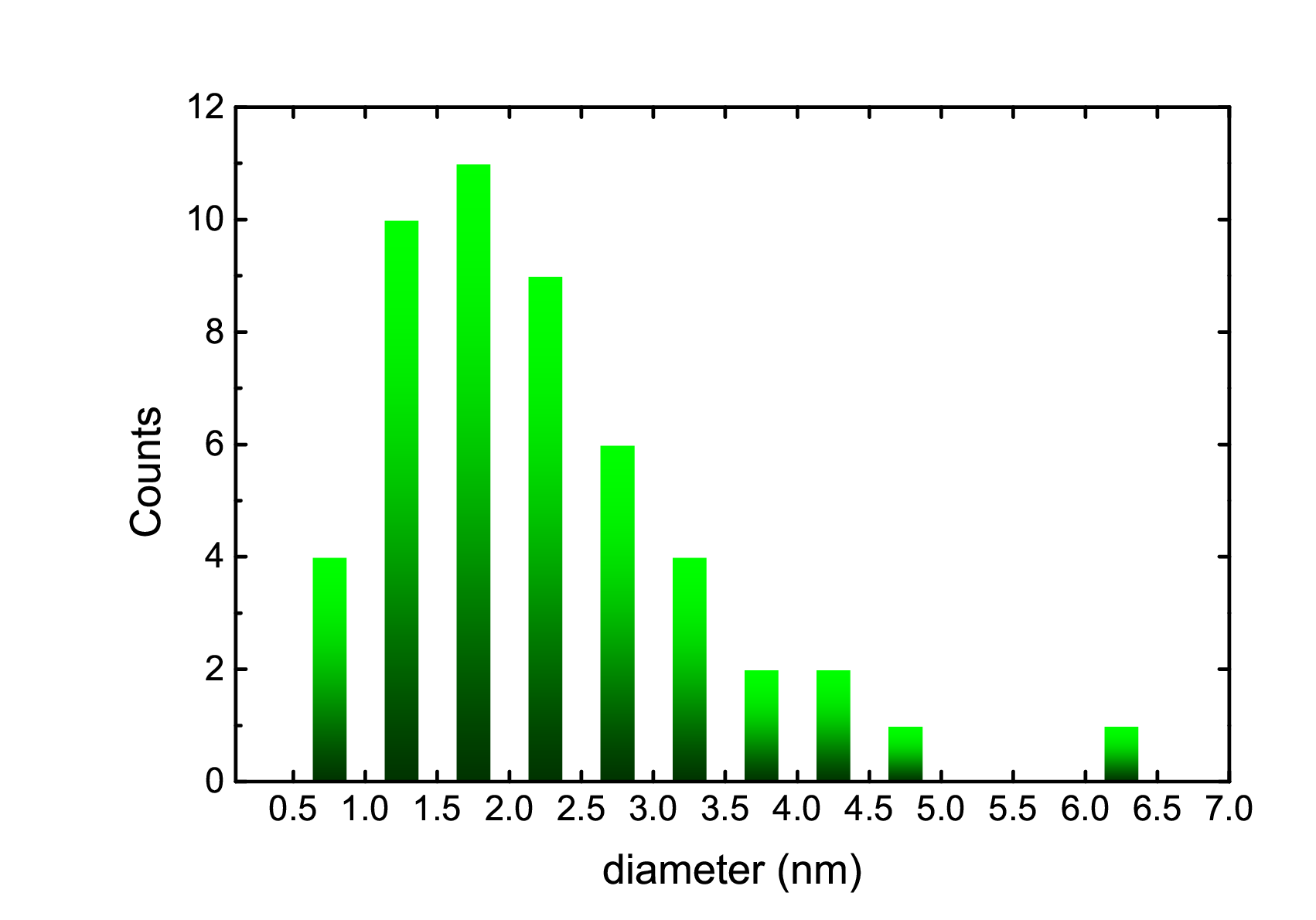}
\caption{\footnotesize Size distribution of the gold nanospheres for which the Mie efficiencies were calculated.}
\label{fig_histogram}
\end{figure}

As possible example to resemble the size distribution of the particles a lognormal distribution was used to calculate a number of 50 samples with randomized diameters fulfilling the lognormal statistics. The mean diameter was set to be 4 nm and the standard deviation was 60$\%$, resulting in the histogram shown in Fig. \ref{fig_histogram}. The Mie efficiencies were calculated independently for each of the nanospheres and, in a second step, averaged for the Mie scattering presented in the manuscript. Thereby, it becomes obvious that in this size regime the particle size distribution has no impact on the resonance position. Thus, the resonance in Fig. 3 in the manuscript remains very sharp and well defined at around 620 nm. In the same manner small deviations from the perfectly spherical shape show only minor impact on the dominant dipole resonance, as carefully investigated in \cite{Granqvist1977}.

Although the resonance position in the imaginary part of the retrieved dielectric function can be predicted by the Mie theory, its very broad width needs to be commented. As pointed out in \citep{Quinten1996} by Quinten and Rostalski in some practical cases the Mie theory might not be sufficient to describe the scattering due absorption of the host medium such as polymers, semiconducting materials, solid dye films or carbon films. Therefore, a more general solution taking into account absorbing host materials was developed (which is not included in the Laven's Mie package). In the case of Ag nanospheres presented by Quinten \cite{Quinten1996}, the width of the resonance was strongly broadened due to the absorbing host. An increase of the absorption index $\kappa$ of the surrounding medium from zero to 0.3 resulted a in 9 times broader resonance. Since k for carbon in the visible and near spectral range can reach values between 0.7-0.9 \cite{Hagemann1975,Schnaiter1998}, the broad resonance appearing in $\text{Im}[\epsilon_{\text{EBID}}]$ can be attributed to the losses of the carbonaceous matrix. The losses of the matrix are also responsible for the small red-shift of the peak position (compared with the Mia calculations presented in manuscript) for the pads of highest gold content, as also reported in \citep{Quinten1996}. 
\subsection{Effective medium approach}
Encouraged by the results, Maxwell-Garnett calculations were performed over a wide range of possible parameters. Since the Maxwell-Garnett theory describes interacting dipoles in a dielectric medium it seems to be ideally suited at the first glance \cite{Kreibig2005}. The particle sizes are by far smaller than $\lambda/10$ in the investigated spectral range and therefore can be safely treated in the quasistatic limit. The observed dipole resonance is almost purely absorptive with a scattering cross-section approximately 100 times smaller than the absorption cross-section (see Fig. 3 in the manuscript). However, varying all the Maxwell-Garnett parameters such as the  filling factor, the refractive index of the non-absorbing host medium and the complex permittivity of embedded gold particles (including size corrections) could not resemble the retrieved dielectric function. 
In this respect it should be mentioned that the Maxwell-Garnett model in its original form does not account for absorptive embedding media \cite{Quinten2011} and the near-field interaction of the closely spaced particles and, thus, fails in many realistic scenarios \cite{Quinten2011,Etruch2014}. Therefore, in case of the EBID composite the theory needs to be adapted appropriately. \\
The first step to tackle these problems is the introduction of self-consistency into the Maxwell-Garnett description \cite{Polder1946}. Accordingly, the background medium can be  subsequently corrected by the effective permittivity itself. Starting with a loss-free dielectric as the host material, the first Maxwell-Garnett calculation provides properties of an effective medium, which in the next iteration is used as the host material. In such a way, one introduces losses iteratively into the host material, which at the beginning is assumed to be purely dielectric. \\
In the next step, this model can be extended by assuming that the gold inclusions can accumulate and form variously shaped aggregates. As it is hard to expect the clusters to be all the same, their presence will alter the size distribution presented in Supplementary Fig.~\ref{fig_histogram}, broadening it and shifting its maximum towards larger sizes. This approach requires the introduction of effective depolarization factors to describe the polarizability of individual aggregates \cite{Granqvist1977, Clippe1976}. The depolarization factors can account for dimers, trimers, linear chains or fcc-type aggregates, which might lead to polarizability anisotropy for different directions and by averaging will constitute another broadening mechanisms. Though, a correction for aggregates might help to analytically understand the EBID composite, it also introduces additional parameters, which need to be experimentally confirmed to correctly model the effective permittivity. \\   
The mentioned methods, the self-consistent approach to the Maxwell-Garnett theory together with the aggregate correction, are intensely investigated by the authors, especially with regard to the up-to-now undefined optical behavior of the carbonaceous matrix. \\

\bibliographystyle{unsrt}
\bibliography{references}